\documentclass[11pt]{article}
\usepackage{amsmath,amssymb,amsfonts,amsthm,dsfont,dcolumn,yfonts}
\usepackage{latexsym}
\usepackage[colorlinks=true, pdfstartview=FitV, linkcolor=black, citecolor=black, urlcolor=black]{hyperref}

\newlength{\textlength}
\newlength{\overlinelength}

\newcommand{\beq}{\begin{equation}}
\newcommand{\eeq}{\end{equation}}
\newcommand{\beqn}{\begin{eqnarray}}
\newcommand{\eeqn}{\end{eqnarray}}
\newcommand\az{{a_z}}
\newcommand\azb{{a_{\bar z}}}
\newcommand\gz{{E_z}}
\newcommand\gzb{{E_{\bar z}}}
\newcommand\hzt{{h^{\bar z}}_t}
\newcommand\hzbt{{h^{z}}_t}

\newcommand{\be}{\begin{equation}}
\newcommand{\ee}{\end{equation}}
\newcommand{\bea}{\begin{eqnarray}}
\newcommand{\eea}{\end{eqnarray}}
\def\ie{{\it i.e.~}}

\def\ads{{\rm AdS}}

\def\G{\Gamma}

\newcommand{\wn}{\textswab{w}}

\begin{document}

\begin{center}\ \\
\vspace{60pt}

{\Large {\bf Transport Coefficients at Zero Temperature\\ from Extremal Black Holes}}\\ 

\vspace{50pt}

{\large Mohammad Edalati, Juan I. Jottar and Robert G. Leigh}
\vspace{10pt}

{\it $^1$Department of Physics, University of Illinois at Urbana-Champaign,\\ Urbana IL 61801, USA}
\\{\tt edalati,jjottar2,rgleigh@illinois.edu}\\ [4mm]

\end{center}

\vspace{70pt}

\centerline{\bf Abstract}

\noindent Using the AdS/CFT correspondence we study transport coefficients of a strongly-coupled $(2 +1)$-dimensional field theory at {\it zero} temperature and finite charge density. The field theory under consideration is dual to the extremal Reissner-Nordstr\"{o}m AdS$_4$ black hole in the bulk. We show that, like the cases of scalar and spinor operators studied in \cite{Faulkner:2009wj}, the correlators of charge (vector) current and energy-momentum (tensor) operators exhibit scaling behavior at low frequency. The existence of such low frequency behavior is related to the fact that the near-horizon geometry of the extremal black hole background has an AdS$_{2}$ factor. We carefully calculate the shear viscosity (at zero temperature) and show that the ratio of the shear viscosity to the entropy density takes the value of  $1/4\pi$.  Because of the AdS$_{2}$ factor, we argue that  this result stays the same for all $d$-dimensional boundary field theories dual to the extremal Reissner-Nordstr\"{o}m AdS$_{d+1}$ black holes. Also, we compute the charge conductivity at zero temperature. The limiting behavior of the conductivity for small frequencies is also attributed to the near horizon AdS$_{2}$ factor and is argued to hold regardless of the dimension of the zero-temperature boundary field theory. Finally, using the extremal dyonic AdS$_4$ black hole as the background, we extract the conductivity in the presence of a constant  magnetic field. 

\newpage

\section{Introduction}
Applying the tools of the AdS/CFT correspondence to study strongly interacting low-dimensional systems with relevance in condensed matter physics is an exciting and  fairly new application of holography.  Since our field theoretic methods are limited when it comes to analyzing strongly interacting systems, the AdS/CFT correspondence enables us to study such systems, or similar ones, as it maps strongly-coupled field theories in $d$ spacetime dimensions to classical gravity in $d+1$ dimensions (See, for example, \cite{Aharony:1999ti} for a review).  Although the theories amenable to analysis using the correspondence are presumably far from realistic condensed matter systems, one may hope that some of the systems studied this way are in some sense of the same universality class. Thus, at a minimum, we may understand some universal features using holography which would have otherwise been impossible.

There are many strongly interacting systems with applications in condensed matter physics which  can be studied using holography.  Following \cite{Gubser:2008px}, a simple gravitational system was used in \cite{Hartnoll:2008vx,Hartnoll:2008kx,Herzog:2008he} to model two of the most important phenomena in condensed matter physics, namely superconductivity and superfluidity. As another application of holography, simple gravity duals have been proposed in \cite{Son:2008ye,Balasubramanian:2008dm,Kachru:2008yh} to model systems with Schr\"{o}dinger and Lifshitz symmetries.    See \cite{Hartnoll:2009sz,Herzog:2009xv} for a review of the recent applications of holography to condensed matter physics

A $(2+1)$-dimensional strongly-coupled field theory at zero temperature and finite U(1) charge density was considered in \cite{Liu:2009dm} which is dual to the background of an extremal (electrically-charged) Reissner-Nordstr\"{o}m black hole with AdS$_4$ asymptotics. Analyzing the spectral function of a fermionic operator, which is  dual to a massless fermion in the background, the authors of {\cite{Liu:2009dm} gave evidence for the existence of Fermi surfaces and observed that, at low energies, near the surface the dispersion relation of the quasi-particle excitations (poles of the spectral function) is of non-Fermi liquid type. This observation along with others was understood in \cite{Faulkner:2009wj}  as stemming from an IR CFT dual to the near horizon region of the black hole background which is ${\rm AdS}_2\times\mathbb{R}^2$. It was shown in \cite{Faulkner:2009wj} that the retarded Green function $G_R(\omega,\vec k)$ can exhibit a scaling behavior for the spectral density, a log-periodic behavior, or  indicate the existence of Fermi surfaces with quasi-particle excitations of non-Fermi liquid type (in the case of fermionic operators). All of these phenomena are attributed to the behavior of certain operators in the IR CFT as well as the behavior of $G_R(0,\vec k)$. Although one expects that the role played by the IR CFT in determining the low energy behavior of the retarded Green functions in the boundary field theory applies to all operators regardless of their nature,  the specific type of phenomena observed may very well depend on whether  $\cal O$ is a scalar, spinor, vector, or a tensor operator. In  \cite{Faulkner:2009wj}, the calculations were carried out for scalar and fermionic operators in $d$-dimensional field theories dual to the extremal Reissner-Nordstr\"{o}m AdS$_{d+1}$ black holes. 

In this paper we consider charge (vector) current and energy-momentum (tensor) operators of the boundary theory at zero temperature and finite charge density and investigate the low energy behavior of their retarded Green functions.  For concreteness, we focus on the case where the boundary theory is $(2+1)$-dimensional. The generalization of our discussions to higher dimensions is straightforward, as the near-horizon geometry of the extremal black holes contains an $\ads_2$ factor in each case. The charge current of the $(2+1)$-dimensional boundary theory corresponds to the electromagnetic perturbations in the transverse directions of the extremal Reissner-Nordstr\"{o}m AdS$_{4}$ black hole while the energy-momentum tensor corresponds to the gravitational perturbations. In this paper we consider those perturbations which are dual holographically to the shear and charge diffusion modes of the boundary theory. 

Taking the small frequency (hydrodynamic) limit at zero temperature is subtle as the $\omega\to 0$ and $T\to 0$ limits do not, in general, commute. So, to calculate transport quantities such as conductivity at zero temperature one should not expect to get the right answer by first considering the system at finite temperature, taking the small frequency limit, then sending the temperature to zero --- $\omega/T$ diverges for any finite $\omega$ in one limit, while it vanishes in the other. In the holographic context, this subtlety arises because the zero temperature (extremal) black hole is more singular at the horizon than for any $T>0$. Understanding how to deal with this near-horizon behaviour  was one of the main technical innovations of \cite{Faulkner:2009wj}. 

Although the temperature is zero in these studies, there is another scale in the problem set by the charge chemical potential $\mu$. Since the modes that we study here do not carry charge, $\mu$ just serves as a scale to which we compare energies and momenta. The study of low frequency correlation functions is then akin to the well-known hydrodynamical calculations at finite temperature.

One of the peculiar features of the charged black hole under consideration is that it has a finite horizon area, even though the temperature vanishes. Apparently this is to be interpreted as a finite ground state degeneracy, which is certainly an odd concept in the dual continuum field theory. It is widely assumed that this is a `large $N$' effect, and $1/N$ corrections would serve to lift the degeneracy. One of the results of this paper is that the shear viscosity, extracted from stress tensor correlators, is proportional to the entropy density, in precisely the same way as it is at finite temperature. 
In fact, we show that this ratio takes the same value of $1/4\pi$ (hence, saturating the KSS bound \cite{Kovtun:2004de}) for all $d$-dimensional boundary field theories (at zero temperature and finite charge density) dual to the extremal Reissner-Nordstr\"{o}m AdS$_{d+1}$ black holes.  We argue that the ratio is the same because for all $d$-dimensional boundary theories dual to the extremal Reissner-Nordstr\"{o}m AdS$_{d+1}$ black holes it can be mapped to the spectral function of a certain operator in the IR CFT. This result is important for two reasons. First, neither the argument for the universality of the ratio based on the so-called ``membrane paradigm"  \cite{Iqbal:2008by} nor the universality argument of \cite{Benincasa:2006fu} applies to black holes at extremality, in which case one has to either generalize the arguments to include the extremal cases or calculate the ratio explicitly for each zero-temperature boundary theory. Second, the boundary theory whose shear viscosity we compute here may be in the same universality class of some quantum critical points which are of interest in condensed matter physics. Typically, these quantum critical points are strongly-coupled CFTs and it is not possible to study them by field theoretic methods. So, it is certainly desirable to calculate some universal properties of these systems. 

The charge conductivity, extracted from vector current-current correlators is also related  to the spectral function of a specific operator in the IR CFT. We show that the longitudinal conductivity of the boundary theory scales as $\omega^2$ in the limit $\omega\to 0$. One should be careful to define the limit in which this calculation pertains, namely $\sigma(\omega/T=\infty,\vec k=0,\omega\to 0^+)$. Furthermore, from the imaginary part of the conductivity we deduce that there is the expected $\delta$-function at $\omega=0$ in the real part of the conductivity. In the final section of the paper, we consider the extremal dyonic black hole corresponding to a boundary theory with a finite electrical charge density and a constant magnetic field. We extract the longitudinal and Hall conductivity in the same limit.

The organization of the paper is as follows. In section \ref{sectiontwo} we introduce the Reissner-Nordstr\"{om} AdS$_{4}$ black hole background and its $\mbox{AdS}_{2}\times \mathds{R}^{2}$ near-horizon region in the extremal case. We then, very briefly, review some of the results in \cite{Faulkner:2009wj} which are of interest to us in this paper: namely, how to carefully treat  the $\omega\to 0$ limit at $T=0$. In section \ref{sectionthree} we consider vector and tensor perturbations of the background and write down the corresponding linearized Einstein-Maxwell equations in radial gauge, confining our attention here to the $\vec k=0$ limit. Section \ref{sectionfour} is devoted to the calculation of the shear viscosity in the boundary field theory. In section \ref{sectionfive} we compute the longitudinal conductivity of the boundary field theory.  Section \ref{sectionsix} extends these ideas to the holographic calculation of the Hall conductivity in the presence of an external magnetic field, by generalizing the background to include magnetic charge. Finally, in section \ref{sectionseven} we discuss our results together with possible extensions and applications for future work.

\section{Reissner-Nordstr\"{o}m AdS$_4$ Background, Boundary Field Theory and IR CFT}\label{sectiontwo}

\subsection{The Background and the Boundary Field Theory}

Consider the Einstein-Maxwell action in 3+1 spacetime dimensions with a negative cosmological constant $\Lambda = -3/L^2$ (where $L$ is the AdS$_4$ radius)
\bea
I = \frac{1}{2\kappa_4^2}\int d^4x \sqrt{-g}\left(R-2\Lambda- L^2F_{\mu\nu}F^{\mu\nu}\right).
\eea
The Reissner-Nordstr\"{o}m AdS$_4$ black hole given by the following metric and gauge field
\begin{align}\label{metric}
ds^2&=g_{\mu\nu}dx^\mu dx^\nu=\frac{r^2}{L^2}\Big(- f(r) dt^2 + dx^2 + dy^2\Big)+\frac{L^2}{r^2 f(r)} dr^2,\\
A&=\mu\Big(1-\frac{r_0}{r}\Big)dt,\label{gfield} 
\end{align}
is a solution to the Einstein-Maxwell equations, where 
\bea\label{fmu}
f(r)=1-M\left(\frac{r_0}{r}\right)^3+Q^2\left(\frac{r_0}{r}\right)^4, \qquad \qquad \mu=\frac{Qr_0}{L^2},
\eea
and the horizon radius $r_0$ is given by the largest real root of $f(r_0)=0$. 
The Hawking temperature of the black hole reads \cite{Chamblin:1999tk}
\bea\label{temp}
T=\frac{r_0}{4\pi L^2}(3-Q^2).
\eea
Also, the entropy, charge and energy densities of the black hole are given by 
\bea\label{ece}
s=\frac{2\pi}{\kappa_4^2}\left(\frac{r_0}{L}\right)^2,\qquad \qquad \rho=\frac{2}{\kappa_4^2}\left(\frac{r_0}{L}\right)^2Q,\qquad\qquad \epsilon=\frac{r_0^3}{\kappa_4^2L^4}M,
\eea
respectively \cite{Chamblin:1999tk, Faulkner:2009wj}. 

The above charged AdS$_4$ background is dual to a $(2+1)$-dimensional strongly-coupled conformal field theory at finite temperature $T$ and finite charge density $\rho$. The entropy and energy densities of the dual theory are given by $s$ and $\epsilon$ in (\ref{ece}), respectively. Also, the asymptotic value of the bulk gauge field $A_t(\infty)=\mu$ is interpreted in the dual theory as the chemical potential for the (electric) charge density.  Little is known about the details of the dual theory from a field theory perspective. On the other hand, using holography much can been learned (especially thermodynamical properties) regarding its strong-coupling behavior; see \cite{Chamblin:1999tk} and its citations.

The expression in (\ref{temp}) indicates that the temperature of the black hole vanishes when $Q^2=3$, \ie the black hole is extremal. In this paper we work in the extremal limit, where the dual theory is at zero temperature but finite charge density. We will refer to this dual theory as the boundary field theory. Since the Einstein-Maxwell equations are invariant under $A_t \to -A_t$, the sign of the chemical potential $\mu$ in the boundary field theory is arbitrary. We take $\mu$ to be positive which, according to (\ref{fmu}), implies that $Q$ is also positive: $Q=\sqrt{3}$.  Although the black hole temperature vanishes at extremality, its horizon area remains finite whose dual interpretation is that  the boundary field theory has a finite entropy density at zero temperature. This is a large $N$ effect.  The gauge fields in supergravity theories are typically coupled to various scalars (moduli) which arise in string theory, or M-theory, compactifications. Taking such couplings into account, the horizon area of extremal black holes should presumably, in generic cases, decrease \cite{Behrndt:1998jd,Myers:2009ij} making eventually the ground state of the corresponding boundary field theories non-degenerate.

\subsection{Near Horizon Region and the IR CFT}

In the extremal limit, $f(r)$ in the background metric (\ref{metric}) takes the form 
\bea\label{extf}
f(r)=1-4\left(\frac{r_0}{r}\right)^3+3\left(\frac{r_0}{r}\right)^4
\eea
which has a double zero at the horizon, and can be approximated near that region (to the leading order in $r-r_0$) by 
\bea
f(r)\simeq \frac{6}{r_0^2}(r-r_0)^2.
\eea
We will often use the scaled (dimensionless) coordinate $u=r/r_0$.
The near horizon geometry is $\ads_2\times \mathbb{R}^2$. To see this, one changes coordinates viz
\beq\label{definition eta}
u=1+\frac{\alpha}{\eta},
\eeq
where\footnote{We include the factor of 6 here and throughout the paper, as it simplifies equations in the near horizon geometry.} $\alpha\equiv L^2/6r_0$. There is then a scaling limit \cite{Faulkner:2009wj} in which
\begin{align}\label{mgnhrads2}
ds^2=\frac{L^2}{6\eta^2}\Big(-dt^2+d\eta^2\Big)+\frac{r_0^2}{L^2}\Big(dx^2+dy^2\Big),\qquad\qquad A&=\frac{Q}{6\eta}dt. 
\end{align}
The curvature radius of the $\ads_2$ is $L/\sqrt{6}$. As usual, the radial coordinate is interpreted holographically as the renormalization scale of the dual field theory, and the near horizon region corresponds to the IR limit. Thus the $\ads_2\times\mathbb{R}^2$ geometry encodes the IR physics ($\omega\to 0$) of the boundary theory.

At least naively, one expects the gravity on the $\ads_2$ space to be dual to a CFT$_1$.  This led the authors of \cite{Faulkner:2009wj} to suggest that the $(2+1)$-dimensional boundary field theory (which is dual to the extremal Reissner-Nordstr\"{o}m $\ads_4$ black hole) flows in the IR to a fixed point described by a CFT$_1$. Following \cite{Faulkner:2009wj} we will refer to this CFT$_1$ as the IR CFT. The details of the $\ads_2/{\rm CFT}_1$ correspondence and how exactly the mapping works are poorly understood. In particular, it is not clear whether one should interpret the theory dual to the $\ads_2$ as a conformal quantum mechanics  or as a chiral  (1+1)-dimensional CFT. 
What is clear from \cite{Faulkner:2009wj} is that whatever this IR theory might be,  it encodes the low frequency behavior of some observables in the full boundary field theory.

Consider a scalar (or a spinor) operator $\cal O$ in the boundary field theory with retarded Green  function $G_R(\omega,\vec k)$. This operator gives rise to a set of operators ${\cal O}_{\vec k}$ in the IR CFT where $\vec k$ is the momentum in $\mathbb{R}^2$. As shown in \cite{Faulkner:2009wj}, the behavior of $G_R(\omega,\vec k)$ at low frequency is mainly governed by the behavior of $G_R(0,\vec k)$ as well as the conformal dimension $\delta_{\vec k}$ of some operators in the IR CFT. In particular, depending on what $G_R(0,\vec k)$ and  $\delta_{\vec k}$ are, $G_R(\omega,\vec k)$ can exhibit a scaling behavior for the spectral density, a log-periodic behavior, or indicate the existence of Fermi surfaces with quasi-particle excitations of non-Fermi liquid type (in the case of $\cal O$ being fermionic). The scaling behavior of the spectral density and its log-periodic behavior are emergent phenomena as they arise from the conformal properties of the IR CFT. 

In what follows, we consider the charge current and the energy-momentum tensor operators of the boundary field theory and elucidate the role of the IR CFT in determining the low frequency behavior of the retarded Green functions of the two operators. More specifically, we focus our attention to the case of zero spatial momenta, $\vec k =0$. This enables us to carefully calculate how the conductivity and the shear viscosity of the zero-temperature boundary field theory depend on various operators in the IR CFT.

\section{Gauge Field and Metric Fluctuations}\label{sectionthree}

The charge current operator in the boundary field theory is dual to the fluctuations of the bulk gauge field in the transverse direction while the energy-momentum tensor is dual to the fluctuations of the metric about the background. The calculation of the corresponding retarded Green functions starts with solving the linearized Einstein-Maxwell equations for these fluctuations. To proceed, define
\bea\label{fluct}
g_{\mu\nu}= \bar g_{\mu\nu}+h_{\mu\nu}, \qquad A_\mu= \bar A_\mu+a_\mu, 
\eea
where $\bar g_{\mu\nu}$ and  $\bar A_\mu$ represent, respectively, the (extremal) metric and the gauge field of the background, and $h_{\mu\nu}$ and $a_\mu$ are the fluctuations. 

In this paper, we shall be concerned with the limit in which we first take $\vec k=0$. As a result, the system retains the $\rm SO(2)$ symmetry in the $xy$-plane, and the various components of the fields can be classified by their $\rm SO(2)$ charge, as well as parity. We thus will consider complex linear combinations of fields. As is familiar from these studies, certain components of the gauge and metric fluctuations will decouple from each other (for example, at finite $\vec k$, one finds that there are two sectors, the `sound' and `shear' channels which differ by parity in the direction transverse to $\vec k$). At $\vec k=0$, the sectors are distinguished by their $\rm SO(2)$ charge as well as parity. In this paper, we will concern ourselves with modes from the charge zero and one sectors in the shear channel, which are involved in the shear viscosity and charge conductivity. The charge one fields are (indices are raised and lowered with the background metric, and we take $z=x+iy$)
\beq
\az=\frac12(a_x-ia_y),\qquad \qquad \hzt=\frac12({h^x}_t-i{h^y}_t).
\eeq
In the case of the charged AdS$_4$ black hole, it is not really necessary to do this, as there is the unbroken $x\leftrightarrow y$ parity symmetry. In a later section, we will study the dyonic black hole (a Reissner-Nordstr\"{o}m AdS$_4$ black hole with both electric and magnetic charges) in which the parity symmetry is broken (the magnetic field chooses an orientation), in which case the complex notation is required.
The charge zero modes of interest are
\beq
{h^z}_z=\frac12\left( {h^x}_x+{h^y}_y-2i{h^x}_y\right),
\eeq
and the conjugate ${h^{\bar z}}_{\bar z}$. One finds that in the radial gauge ${h^x}_x+{h^y}_y\propto \sqrt{f(r)}$ is constrained, so the remaining propagating field is ${h^x}_y$, dual to the operator $T_{xy}$ of the boundary theory. The other charge zero fields $a_t$ and ${h^t}_t$ have opposite parity and decouple from ${h^x}_y$.

The linearized Einstein-Maxwell equations  then read
\begin{align}
&u^2f(u){h^x}_y''(u)+ \left[u f'(u)+4f(u)\right]u{h^x}_y'(u)+\frac{(6\alpha)^2\omega^2}{u^2f(u)}{h^x}_y(u)=0,\label{shearxy}\\ 
&u^2 f(u) \az''(u)+ \left[uf'(u)+2f(u)\right]u\az'(u)+\frac{Q}{6\alpha}{\hzt}'(u)+\frac{(6\alpha)^2\omega^2}{u^2f(u)}a_z(u) =0,\label{eqy}\\ 
&u^4\hzt''(u)+ 4u^3\hzt'(u)+24\alpha Q\az'(u)=0, \label{eqyt}
\end{align}
where $f(u)$ is given in (\ref{extf}) and the primes denote derivative with respect to $u$.  Note that  $Q=\sqrt{3}$. There is a constraint which comes from the $zu$-component of the linearized Einstein equations
\bea\label{cons}
u^4 \hzt'(u)+24\alpha Q\az(u)=0
\eea
which is clearly the first integral of (\ref{eqyt}). 
Therefore the above system of equations reduces to a single second order equation for $\az$ (the same equation is satisfied by both $a_x$ and $a_y$ in this case) and a second order equation for ${h^x}_y$. 
In what follows, we will often use the notation $ \wn=\alpha\omega.$ Thus frequencies are compared to the scale $\alpha^{-1}\equiv 6r_0/L^2$, which by virtue of the relation $\alpha=Q/6\mu$, is essentially the chemical potential, the only scale of the zero temperature boundary theory.

\section{Shear Viscosity of the Boundary Field Theory}\label{sectionfour}

There is a Kubo formula which relates the shear viscosity, denoted\footnote{Throughout the paper we denote by $\eta$ both the radial coordinate of the near horizon $\ads_2\times\mathbb{R}^2$ geometry and the shear viscosity of the boundary field theory. The distinction between the two is clear from the context.} by $\eta$,  to the retarded Green function of the energy-momentum tensor $T_{xy}$ at $\vec k=0$:  
\bea\label{shearkubo}
\eta=-\lim_{\omega\to 0} \frac{1}{\omega}{\rm Im}~G^R_{xy,xy}(\omega,0).
\eea
To calculate $G^R_{xy,xy}(\omega,0)$ holographically, one solves the equation of motion for ${h^x}_y(u)$ subject to the in-falling boundary condition at the horizon \cite{Son:2002sd, Policastro:2002se}. Note that equation (\ref{shearxy}) for ${h^x}_y(u)$ is identical to the equation of motion for a minimally-coupled chargeless massless  scalar $\phi(u)$ in the charged $\ads_4$ background (\ref{metric}) at extremality. 

Before we calculate $G^R_{xy,xy}(\omega,0)$, we should remind the reader that the zero-temperature boundary field theory has a finite entropy. The fact that there is a finite density at zero temperature makes one wonder about the existence of hydrodynamical modes, such as shear, in the same way that there exist such modes at finite temperature. Suppose the (2+1)-dimensional zero-temperature boundary field theory that we are considering has a non-trivial shear viscosity. As $\mu$ is the only scale in the boundary field theory, one expects, on general grounds, that $\eta=g(N,\lambda) \mu^2$, where $N$ is the number of species and $\lambda$ is the effective (dimensionless) 't Hooft coupling. The AdS/CFT correspondence relates $G^R_{xy,xy}(\omega,0)$ to the on-shell action for ${h^x}_y$, which has no $\lambda$ dependence to the leading order. This implies that $g(N,\lambda)$ is not only independent of $\lambda$ but also proportional to $\kappa_4^{-2}$. Since $g(N)$ is dimensionless one deduces that $g(N)\propto L^2/\kappa_4^2$. Note that $r_0$ cannot be used to make $g(N)$ dimensionless as in the AdS/CFT correspondence, independent of the nature of the gravity solution, the parameter $N$ to some power is mapped to the ratio of the AdS radius to the Planck length.\footnote{Indeed, if the boundary field theory can be realized in some regime as the world-volume theory of $N$ M2-branes, then $N^{3/2} \propto L^2/\kappa_4^2$ \cite{Itzhaki:1998dd, Herzog:2002fn}.} Putting pieces together, we conclude that $\eta \propto L^2\mu^2/\kappa_4^2$, or by virtue of (\ref{ece}), $\eta\propto s$.

At finite temperature, there are general arguments \cite{Iqbal:2008by, Benincasa:2006fu} stating that at large $N$ and large $\lambda$, the ratio of the shear viscosity to the entropy density of the holographic dual field theories (at finite temperature and charge density) equals $1/4\pi$. These arguments assume that the black hole background has a finite temperature, \ie $f(u)$ has a single zero at the horizon. What makes the calculation of shear viscosity of the zero-temperature boundary field theory (which has a finite entropy) non-trivial is the fact that the $\omega\to 0$ and $T\to 0$ limits do not, in general, commute. When the black hole background is extremal, extra care (to be explained below) is required in taking the $\omega\to 0$ limit of ${\rm Im}~G^R_{xy,xy}(\omega,0)$. 

For the ease of notation we take ${h^x}_y\equiv \phi$. Since we will be interested in the small $\omega$ limit, we expand fields in a power series
\beq\label{outerexpan}
\phi_O(u)=\phi^{(0)}_O(u)+\wn~\phi^{(1)}_O(u)+\wn^2 \phi^{(2)}_O(u)+\ldots~.
\eeq
Here,  the subscript  denotes that the expansion is in terms of the $u$ coordinate, which is a suitable coordinate in the asymptotic region (outer region). The $\omega$ expansion near the horizon is subtle because $f(u)$ has a double zero there, and a proper treatment was explained in \cite{Faulkner:2009wj}. One first notes that if the full geometry is replaced by the near-horizon geometry, then the equations of motion organize themselves as functions of $\zeta\equiv \omega\eta$, with no further $\omega$ dependence. We thus have
\beq\label{inner}
u=1+\frac{\wn}{\zeta}.
\eeq
 For example, the equation for ${h^x}_y$ in the near horizon geometry is simply
\bea
\Big(\frac{d^2}{d\zeta^2}+1\Big){h^x}_y(\zeta)=0.
\eea
Since the strict near horizon geometry is to be interpreted as the IR limit, the $\omega$ expansion near the horizon is then properly organized as a series
\begin{align}
\phi_I(\zeta)&=\phi^{(0)}_I(\zeta)+\wn~\phi^{(1)}_I(\zeta)+\wn^2 \phi^{(2)}_I(\zeta)+\ldots\label{innerexpan}
\end{align}
with the leading term gotten by studying the problem in the near-horizon $\ads_2\times\mathbb{R}^2$ geometry. The subscript  $I$ denotes that the expansion (\ref{innerexpan}) is in terms of the $\zeta$ coordinate, the suitable radial coordinate for the $\ads_2$ part (inner region) of the near horizon geometry. The boundary condition is applied at the horizon in the $\zeta$ coordinate, and the solution must be matched on to the outer solution. This is accomplished by taking a double limit, $\zeta\to 0$ and $\wn/\zeta\to 0$. Indeed, we just need to match  $\phi^{(0)}_I(\zeta)$  to $\phi^{(0)}_O(u)$ near the matching region where we take the aforementioned double limit. This is because the differential equation (\ref{shearxy}) is linear,  and we also require that the solutions for the higher order terms in the expansions (\ref{outerexpan}) and (\ref{innerexpan}) do not include terms proportional to  the solution of $\phi^{(0)}_O(u)$, or $\phi^{(0)}_I(\zeta)$, near the matching region \cite{Faulkner:2009wj}.  This is also the case when we match the solutions (for the gauge field fluctuations) in the later sections.

\subsection{Inner Region, and a Dimension 1 Operator in the IR CFT}

In the cases that we will study in this paper, the fluctuations (being abelian gauge and metric modes) are not charged under the U(1) background gauge field. Consequently, in our case small $\wn$ means that $\wn\ll1$. 

Expressed in terms of the coordinate $\zeta$, equation (\ref{shearxy}) becomes
\bea\label{shearz}
\Big(1+\frac{2\wn}{3\zeta}+\frac{\wn^2}{6\zeta^2}\Big)\zeta^2\phi''(\zeta)-\frac{2\wn}{3\zeta}\Big(1+\frac{\wn}{2\zeta}\Big)\zeta\phi'(\zeta)+\Big(1+\frac{\wn}{\zeta}\Big)^4\Big(1+\frac{2\wn}{3\zeta}+\frac{\wn^2}{6\zeta^2}\Big)^{-1}\zeta^2\phi(\zeta)=0,
\eea
where the primes now denote derivatives with respect to $\zeta$. Notice that taking the $\wn\to 0$ limit of (\ref{shearz}) is now smooth near the horizon.
Since the goal is to compute a retarded Green function (for small $\wn$) in the boundary field theory, we solve the inner region equation with in-falling boundary condition for $\phi_I (\zeta)$ at the horizon \cite{Son:2002sd}. 

To the leading order, equation (\ref{shearz}), which becomes an equation for $\phi^{(0)}_I(\zeta)$, reads 
\bea\label{ieq0}
\phi^{(0)\prime\prime}_I(\zeta)+\phi^{(0)}_I(\zeta)=0.
\eea
We solve (\ref{ieq0}) with in-falling boundary condition at the horizon. The general solution of (\ref{ieq0}) is
\bea\label{generalinnersol}
\phi^{(0)}_I(\zeta)= a^{(0)}_I e^{i\zeta}+b^{(0)}_I e^{-i\zeta}.
\eea
Choosing the in-falling boundary condition at the horizon means that we need to discard the $e^{-i\zeta}$ branch of the solution by setting $b^{(0)}_I=0$. Expanding the result near the matching region (boundary of $\ads_2$), we obtain
\bea\label{bisol0}
\phi^{(0)}_I(\zeta)|_{\zeta\to 0}\simeq a^{(0)}_I \left(1+i\zeta\right)=a^{(0)}_I \left[1+{\cal G}_R(\wn)\frac{1}{u-1}\right],
\eea
where we have used (\ref{inner}) to express $\zeta$ in terms of $u$, and 
\bea\label{greenonehalf}
{\cal G}_R(\wn)=i\wn
\eea
is the retarded Green function of a chargeless scalar operator $\cal O$ with a conformal dimension of $\delta=1$ in the IR CFT.\footnote{
Generally, the Green function of an operator of dimension $\delta=\nu+\frac{1}{2}$ in the IR CFT is given by 
\begin{align}\label{tworetgreen}
{\cal G}_R(\omega)=-2\nu~e^{-i\pi\nu}\frac{\G(1-\nu)}{\G(1+\nu)}\left(\frac{\omega}{2}\right)^{2\nu}
\end{align}
which is a simplification of a formula given in \cite{Faulkner:2009wj}.}  The relationship between $\phi^{(0)}_I$ and ${\cal G}_R(\wn)$ in (\ref{bisol0}) can be easily seen by noticing \cite{Faulkner:2009wj} that, once $\zeta$ is rescaled by a factor of $\wn$ (namely, $\zeta\to\zeta/\wn$),  equation (\ref{ieq0}) becomes identical to the equation of motion (in Fourier space) for a massless, chargeless scalar field in $\ads_2$. Based on general grounds of the AdS/CFT correspondence, $\phi^{(0)}_I$ is dual to the aforementioned scalar operator $\cal O$ in the IR CFT.

It turns out that for the purpose of calculating the shear viscosity of the boundary field theory using the Kubo formula (\ref{shearkubo}), it suffices to just know (\ref{bisol0}). Nevertheless, solving equation (\ref{shearz}) perturbatively in $\wn$, it could be easily verified that the inner region solution takes the following form near the matching region (keeping in mind that $\zeta\to0$ and $\wn/\zeta\to 0$ in that region)
\bea\label{innerexpansol}
\phi_I(u)=a^{(0)}_I\left\{\left[1+\ldots\right]+\frac{1}{u-1}{\cal G}_R(\wn)[1+\ldots]\right\},
\eea
where the dots represent terms which vanish in the $u\to 1$ and $\wn\to 0$ limits.

\subsection{Outer Region}

As we will see below, to calculate the shear viscosity of the boundary field theory, it suffices just to find the solution for the zeroth-order term in (\ref{outerexpan}), namely $\phi^{(0)}_O(u)$. Substituting (\ref{outerexpan}) into (\ref{shearxy}),  $\phi^{(0)}_O(u)$ obeys the following equation
\bea\label{zerothouter}
u^4f(u)\phi_O^{(0)\prime\prime}(u)+ u^3\left[4f(u)+ r f'(u)\right]\phi_O^{(0)\prime}(u)=0,
\eea
whose solution reads
\bea\label{solphi0}
\phi^{(0)}_O(u)=a^{(0)}_O+\frac{1}{36}b^{(0)}_O\Big[-\frac{6}{u-1}-4~{\rm ln}(u-1)+\sqrt{2}~{\rm tan}^{-1}\Big(\frac{u+1}{\sqrt{2}} \Big)+2~{\rm ln}(u^2+2u+3)\Big] .
\eea
Near the boundary $u\to \infty$, (\ref{solphi0}) takes the form
\bea\label{outerfullboundary}
\phi^{(0)}_O(u)|_{u\to\infty}=\Big(a^{(0)}_O+\frac{\pi}{36\sqrt{2}}b^{(0)}_O\Big)-\frac{b^{(0)}_O}{3u^3}+\ldots,
 \eea
where the dots represent terms subleading compared to $u^{-3}$. Near the matching region, (\ref{solphi0}) becomes
\bea\label{solzerothmatch}
\phi^{(0)}_O(u)|_{u\to 1}&=&-\frac{b^{(0)}_O}{6(u-1)}\Big\{1+\ldots\Big\}+\Big[a^{(0)}_O+\frac{b^{(0)}_O}{36}\Big(\sqrt{2}~{\rm tan}^{-1}(\sqrt{2})+{\rm ln}(36)\Big)\Big]\Big\{1+\ldots\Big\},
\eea
with the dots representing terms which vanish as $u\to 1$ and $\wn\to 0$.

\subsection{Matching, and the Shear Viscosity}

Matching (\ref{innerexpansol}) with (\ref{solzerothmatch}) results in
\bea\label{coeff}
a^{(0)}_O=\Big[1+\frac{1}{6}{\cal G}_R(\wn)\Big(\sqrt{2}~{\rm tan}^{-1}(\sqrt{2})+{\rm ln}(36)\Big)\Big]a^{(0)}_I, \qquad b^{(0)}_O=-6{\cal G}_R(\wn) a^{(0)}_I.
\eea
Knowing the above two coefficients (in terms of $a^{(0)}_I$) is enough to compute the shear viscosity of the boundary field theory.  Substituting (\ref{coeff}) into (\ref{outerfullboundary}), we obtain 
\bea\label{asymfinalsol}
\phi_O(u)|_{u\to \infty}=a^{(0)}_I[1+s_0 {\cal G}_R (\wn)+\ldots]+2a^{(0)}_I{\cal G}_R(\wn) [1+\ldots] u^{-3}+{\cal O}\left(u^{-6}\right),
\eea
where $s_0$ is a numerical factor given by
\bea
s_0=\frac{1}{6\sqrt{2}}\Big(2~{\rm tan}^{-1}(\sqrt{2})+\sqrt{2}~{\rm ln}(36)-\pi\Big).
\eea

Recall that $\phi(u)$ denotes ${h^x}_y(u)$. Thus, (\ref{asymfinalsol}) gives the asymptotic form of  ${h^x}_y(u)$ near the boundary of the extremal charged $\ads_4$ black hole. Reading off the normalization of ${h^x}_y(u)$ from the Einstein-Hilbert action (plus the Gibons-Hawking term) and applying the real-time recipe  of \cite{Son:2002sd},  we have 
\bea\label{greenboundary}
G^R_{xy,xy}(\omega,0)=-\frac{1}{2\kappa_4^2}\left(\frac{r_0}{L}\right)^2{\cal G}_R(\omega)\left[1+{\cal O} (\omega)\right].
\eea
Note that in (\ref{greenboundary}),  ${\cal G}_R(\omega)=i\omega$. The Kubo formula (\ref{shearkubo}) then yields
\bea\label{threeviscosity}
\eta= \frac{L^2}{2\kappa_4^2}\left(\frac{\mu}{Q}\right)^2\Big[\lim_{\omega\to 0}\frac{1}{\omega}{\rm Im}~{\cal G}_R(\omega)\Big].
\eea
Rather than substituting its value, we kept ${\rm Im}~{\cal G}_R(\omega)$ explicit in the above formula to emphasize that the shear viscosity of the boundary field theory is given by the spectral function of an IR CFT scalar operator whose conformal dimension is $\delta=1$.

Using (\ref{threeviscosity}) and the formula in (\ref{temp}) for the entropy density, one then deduces that     
\bea\label{etasratio}
\frac{\eta}{s}=\frac{1}{4\pi}
\eea
in our (2+1)-dimensional boundary field theory which is at zero temperature and finite charge density. Note that the ratio of the shear viscosity to the entropy density given in (\ref{etasratio}) has been obtained by properly taking care of the $\omega\to 0$ limit at $T=0$. Our explicit calculation above confirms that $\eta/s$ takes the same value of $1/4\pi$ in the zero-temperature boundary field theory as it does in the finite temperature boundary field theory. One may then assume that the zero-temperature result for $\eta/s$ can be obtained by an extrapolation from the finite-temperature result. This assumption is not true as the $\omega\to 0$ and $T\to 0$ limits do not in general commute.  

We should add here that the (2+1)-dimensional boundary field theory whose shear viscosity we have just computed  may be in the same universality class of some quantum critical points; see \cite{Faulkner:2009wj} for a discussion in this direction.  Often these quantum critical points are strongly-coupled and we do not have a satisfactory field theoretic description. Shear viscosity is one of the quantities one would like to calculate for these systems (in the quantum critical region).

Incidentally, the same calculations can be carried out in other dimensions as well. In the extremal limit, the near-horizon geometry contains an $\ads_2$ factor, and, as we have checked explicitly, the shear viscosity is determined by the same $\delta=1$ operator. Consequently, $\eta/s=1/4\pi$ in each of these cases. 

\section{Conductivity of the Boundary Field Theory}\label{sectionfive}

In this section, we consider the charge conductivity of the zero temperature, finite charge density boundary theory for $\omega\to 0$. Because of the symmetry, we will discuss one real component of the gauge field, and the other component (or either of the complex linear combinations) follows in the same way. The fluctuation $a_y$ is dual to a conserved current $J_y$ in the boundary field theory. There is a  Kubo formula which relates the conductivity $\sigma_{yy}$ to the retarded Green function $G^R_{yy}(\omega, k)$ of $J_y$ at $\vec k=0$:
\bea\label{condkubo}
\sigma_{yy}=\lim_{\omega\to 0^+}\frac{i}{\omega} G^R_{yy}(\omega,0).
\eea

From now on, we denote $\sigma_{yy}$ simply by $\sigma$. Note that $\sigma$ is dimensionless in (2+1) dimensions. To calculate $\sigma$, we need to solve the linearized Maxwell equation for $a_y$ at $\vec k=0$. Combining equations  (\ref{eqy}) and (\ref{cons}) we find
\bea\label{ay}
u^2 \Big[u^2f(u)a'_y(u)\Big]'+ 12\left(3\frac{\wn^2}{f(u)}-\frac{1}{u^2}\right)a_y(u)=0.
\eea
Similar to the previous problem, the $\wn\to 0$ limit  becomes subtle. Again we employ two expansions
\begin{align}\label{condouterexpan}
a_{yO}(u)&=a^{(0)}_{yO}(u)+\wn a^{(1)}_{yO}(u)+\wn^2 a^{(2)}_{yO}(u)+\ldots,\\
a_{yI}(\zeta)&=
a^{(0)}_{yI}(\zeta)+\wn a^{(1)}_{yI}(\zeta)+\wn^2 a^{(2)}_{yI}(\zeta)+\ldots, \label{condinnerexpan}
\end{align}
where $a^{(0)}_{yI}(\zeta)$ is the solution in the near horizon geometry.

\subsection{Inner Region}

Written in terms of the coordinate $\zeta$, equation (\ref{ay}) takes the form
\begin{align}\label{coninnereqn}
&\Big(1+\frac{2\wn}{3\zeta}+\frac{\wn^2}{6\zeta^2}\Big)^2\Big(1+\frac{\wn}{\zeta}\Big)^2\zeta^2a''_y(\zeta)+\frac{4\wn}{3\zeta}\Big(1+\frac{5\wn}{4\zeta}+\frac{\wn^2}{4\zeta^2}\Big)\Big(1+\frac{2\wn}{3\zeta}+\frac{\wn^2}{6\zeta^2}\Big)\zeta a'_y(\zeta)\nonumber\\
&+\left[\zeta^2\Big(1+\frac{\wn}{\zeta}\Big)^6-2\Big(1+\frac{2\wn}{3\zeta}+\frac{\wn^2}{6\zeta^2}\Big)\right]a_y(\zeta)=0.
\end{align}
To the leading order, we have the equation of motion in the $\ads_2$ geometry
\bea\label{icondzerotheq}
a^{(0)\prime\prime}_{yI}(\zeta)+ \Big(1-\frac{2}{\zeta^2}\Big)a^{(0)}_{yI}(\zeta)=0.
\eea
The general solution reads 
\bea\label{icondrawzsol}
a^{(0)}_{yI}(\zeta)= a^{(0)}_I \Big(1+\frac{i}{ \zeta}\Big)e^{i\zeta}+b^{(0)}_I \Big(1-\frac{i}{\zeta}\Big) e^{-i\zeta}.
\eea
The second term in (\ref{icondrawzsol}) gives rise to an out-going wave at the horizon. Thus,  we set  $b^{(0)}_I=0$. Near the matching region, (\ref{icondrawzsol}}) becomes
\begin{align}\label{simpicondzsol}
a^{(0)}_{yI}(\zeta)|_{\zeta\to 0}= a^{(0)}_I \left\{\frac{i}{\zeta}[1+\ldots]-\frac{\zeta^2}{3}[1+\ldots]\right\},
\end{align}
where the dots represent terms which vanish as $\zeta\to 0$ and $\wn/\zeta \to 0$. Rescaling $a^{(0)}_I\to -i\wn a^{(0)}_I$, recalling the definition of $\zeta$ in terms of $u$, and dropping the subleading terms, (\ref{simpicondzsol}) takes the form
\bea\label{greenicondzsol}
a^{(0)}_{yI}(u)|_{u\to 1}\simeq a^{(0)}_I\left\{(u-1)+{\cal G}_R(\wn)\frac{1}{3(u-1)^2}\right\},
\eea
where 
\bea\label{ircftgreentwo}
{\cal G}_R(\wn)=i \wn^3
\eea
is the retarded Green function of a (chargeless) scalar operator $\cal O$ in the IR CFT whose conformal dimension is $\delta=2$; see footnote 4 for details. The $\wn$ expansion of the inner solution can be continued to higher orders but the result above is sufficient for our purposes.
The terms $a^{(n>0)}_{yI}(\zeta)$ are irrelevant  to calculating $\sigma$ at low frequency  using the Kubo formula (\ref{condkubo}).

\subsection{Outer Region}

To calculate $\sigma$, it suffices to solve for $a^{(0)}_{yO}(u)$ in the outer region expansion (\ref{condouterexpan}). To the leading order, (\ref{ay}) gives
\bea\label{ocondzeroth}
u^4\left[u^2f(u)a^{(0)\prime}_{yO}(u)\right]'-12a^{(0)}_{yO}(u)=0
\eea
whose solution is found to be
\begin{align}\label{fullocondzeroth}
a^{(0)}_{yO}&=a^{(0)}_{O}\Big(1-\frac{1}{u}\Big)+ b^{(0)}_{O}\Big\{-\frac{65u^2-100u+41}{108u(u-1)^2}\nonumber\\
&+\Big(1-\frac{1}{u}\Big)\Big[\frac{17\sqrt{2}}{648}{\rm tan}^{-1}\Big(\frac{u+1}{\sqrt{2}}\Big)+\frac{14}{81}{\rm ln}(u-1)-\frac{7}{81}{\rm ln}(u^2+2u+3)
\Big]\Big\}.
\end{align}
The solution (\ref{fullocondzeroth}) near the boundary $u\to \infty$ takes the form 
\bea
a^{(0)}_{yO}(u)|_{u\to\infty}=\Big(a^{(0)}_{O}+\frac{17\pi}{648\sqrt{2}}b^{(0)}_{O}\Big)-\Big[a^{(0)}_{O}+\Big(1+\frac{17\pi}{648\sqrt{2}}\Big)b^{(0)}_{O}\Big]\frac{1}{u}+{\cal O}\left(u^{-2}\right),
\eea
whereas near the matching region, it becomes
\begin{align}
a^{(0)}_{yO}(u)|_{u\to 1}= &-\frac{b^{(0)}_{O}}{18(u-1)^2}\left[1+{\cal O}(u-1)\right]\nonumber\\
&+\Big[a^{(0)}_{O}+\frac{b^{(0)}_{O}}{648}\left(246+17\sqrt{2}~{\rm tan}^{-1}(\sqrt{2})-56~{\rm ln}(6)\right)\Big](u-1)\left[1+{\cal O}(u-1)\right].
\end{align}
The asymptotic forms of $a_{yO}(u)$ thus read
\begin{align}
a_{yO}(u)|_{u\to\infty}&\simeq\Big(a^{(0)}_{O}+\frac{17\pi}{648\sqrt{2}}b^{(0)}_{O}\Big)\Big\{1+\ldots\Big\}-\Big[a^{(0)}_{O}+\Big(1+\frac{17\pi}{648\sqrt{2}}\Big)b^{(0)}_{O}\Big]\Big\{1+\ldots\Big\}\frac{1}{u},\label{condboundregion}\\ 
a_{yO}(u)|_{u\to 1}&\simeq -\frac{b^{(0)}_{O}}{18(u-1)^2}\Big\{1+\ldots\Big\}\nonumber\\
&+\Big[a^{(0)}_{O}+\frac{b^{(0)}_{O}}{648}\left(246+17\sqrt{2}~{\rm tan}^{-1}(\sqrt{2})-56~{\rm ln}(6)\right)\Big]\Big\{1+\ldots\Big\}(u-1).\label{condoutermatchregion}
\end{align}
The dots in (\ref{condboundregion}) and (\ref{condoutermatchregion}) represent terms which vanish as $\wn\to 0$.

\subsection{Matching and the Conductivity}

Matching (\ref{greenicondzsol}) to (\ref{condoutermatchregion}), we obtain
\bea\label{condcoeff}
a^{(0)}_O=\left[1+\frac{{\cal G}_R(\wn)}{108}\left(246+17\sqrt{2}~{\rm tan}^{-1}(\sqrt{2})-56~{\rm ln}(6)\right)\right]a^{(0)}_I, \qquad b^{(0)}_O=-6{\cal G}_R(\wn) a^{(0)}_I.
\eea
Plugging (\ref{condcoeff}) into (\ref{condboundregion}) yields
\bea\label{finalcondasymp}
a^{(0)}_{yO}(u)|_{u\to\infty}\simeq a^{(0)}_I \Big[1+s_1{\cal G}_R(\wn)+\ldots\Big]-a^{(0)}_I \Big[1+s_2{\cal G}_R(\wn)+\ldots\Big]\frac{1}{u}.
\eea
where
\bea
s_1=\frac{1}{108\sqrt{2}}\left(246\sqrt{2}+34~{\rm tan}^{-1}(\sqrt{2})-56\sqrt{2}~{\rm ln}(6)-17\pi\right),\qquad s_2=s_1-6.
\eea
From (\ref{finalcondasymp}) and the normalization of the Maxwell action for $a_y(u)$, the leading small frequency behavior of $G^R_{yy}(\omega, 0)$ is found to be 
\begin{align}
{\rm Im}~G^R_{yy}(\omega, 0)=-\frac{2\alpha^2 L^2}{\kappa_4^2}~ {\rm Im}~{\cal G}_R (\omega),
\end{align}
where ${\rm Im}~{\cal G}_R (\omega)=\omega^3$. The Kubo formula (\ref{condkubo}) then gives  
\bea\label{threeconductivity}
{\rm Re}~ \sigma(\omega\to 0)=\frac{2\alpha^2 L^2}{\kappa_4^2}\lim_{\omega\to 0} \frac{1}{\omega}{\rm Im}~{\cal G}_R(\omega)=\frac{L^2}{6\kappa_4^2} \lim_{\omega\to 0} \left(\frac{\omega}{\mu}\right)^2=0.
\eea
Thus, we see that a dimension 2 scalar operator in the IR CFT determines the limiting behavior as $\omega\to 0$ of the conductivity of the (2+1)-dimensional zero-temperature boundary field theory. In addition\footnote{We thank J. McGreevy for pointing out that we omitted a discussion of ${\rm Im}~\sigma$ in an earlier version of this paper.}, the imaginary part of $\sigma$ has a pole at $\omega=0$
\beq
{\rm Im}~\sigma (\omega\to 0) = \frac{\alpha^2 L^2}{3\kappa_4^2}\frac{1}{\omega}+...
\eeq
A Kramers-Kronig relation (see for example, Ref. \cite{Hartnoll:2008kx}) then implies that the real part has a delta function at $\omega=0$ (which is not detected directly by the limiting procedure for ${\rm Re}~\sigma$ described above)
\beq\label{realdelta}
{\rm Re}~\sigma= \frac{\pi L^2}{36\kappa_4^2\mu^2}\delta(\omega).
\eeq

This result should be contrasted with the conductivity of the boundary theory at finite temperature. For strongly-coupled (2+1)-dimensional theories at finite temperature, the AdS/CFT correspondence gives a constant temperature-independent result for the conductivity \cite{Kovtun:2008kx, Iqbal:2008by} at zero density, whose calculation requires taking the $\frac{\omega}{T}\to 0$ limit. On the other hand, what we computed above is effectively the behavior of the conductivity in the $\frac{\omega}{T}\to \infty$ limit (as well as in the $\frac{\omega}{\mu}\to 0$ limit) had we heated up the zero-temperature boundary field theory. Contrary to the calculation of the shear viscosity, different results are obtained for the conductivity in the two different limits of  $\frac{\omega}{T}\to 0$ and $\frac{\omega}{T}\to \infty$.  This indicates that, in general, one does not expect the $\omega\to 0$ and $T\to 0$ limits to commute. The subtlety in the order of limits and its origin has been discussed in \cite{Herzog:2007ij}. Our result that the limiting value of the conductivity of the zero-temperature boundary field theory under consideration is zero relies heavily on the fact that the theory is at finite charge density (which, in turn, is intertwined with the existence of the $\ads_2$ region in the background geometry).  In the case of zero charge density, the authors of \cite{Herzog:2007ij} have given a general argument that the conductivity should be constant for all values of $\frac{\omega}{T}$. 

For the $d$-dimensional generalizations of the boundary field theory we considered above, the same dimension 2 scalar operator in the IR CFT enters the computation of the conductivity of the zero-temperature boundary field theory, as we have checked explicitly. Thus, to the leading order, ${\rm Im}\ G^R_{yy}(\omega\to 0, 0)$ scales as $\omega^2$. These results along with (\ref{threeconductivity}) give enough evidence that the conductivity of the $d$-dimensional zero-temperature boundary field theories at finite charge density, dual to the extremal charged $\ads_{d+1}$ black holes, behaves as 
\bea\label{generalcon}
\sigma(\omega\to 0)\sim \frac{L^{3-d}r_0^{d-2}}{\kappa^2_{d+1}~\mu^3}\lim_{\omega\to 0} \frac{1}{\omega}{\rm Im}~{\cal G}_R(\omega)\sim\frac{L^{5-d}r_0^{d-3}}{\kappa^2_{d+1}}\lim_{\omega\to 0} \left(\frac{\omega}{\mu}\right)^2,
\eea
where $L$ and $r_0$ in (\ref{generalcon}) denote the curvature radius and the horizon of the extremal charged $\ads_{d+1}$ black holes, respectively. The imaginary part contains the same pole at $\omega=0$, and hence there is a similar peak in ${\rm Re}~\sigma$ at $\omega=0$, as in eq. (\ref{realdelta}). We expect that if this peak is to be avoided, the spatial translation invariance of the system must be broken.

\section{Hall Conductivity}\label{sectionsix}

The calculation of the Hall conductivity for the boundary field theory at  finite temperature was considered, for example,  in \cite{Hartnoll:2007ai, Hartnoll:2007ip}. In this section we calculate the conductivity of the $(2+1)$-dimensional zero-temperature boundary field theory in the presence of a background magnetic field.

The Hall conductivity at $\omega\to 0$ is an example of a quantity that is {\it not} determined by properties of the IR CFT. This is to be expected because, as is well known, it follows from simple symmetry arguments. We have included the calculation in this paper to demonstrate that indeed the zero temperature limit can be taken, following the methods of \cite{Faulkner:2009wj}. This is probably the simplest example in which these methods can be applied when fluctuations are mixing, (this also happens, for example, generically at finite momentum). We will find evidence below that additional subtleties arise in the non-commutavity of the $\omega\to 0$ and $H\to 0$ limits.

To consider the $(2+1)$-dimensional  boundary field theory in a background magnetic field, one adds a magnetic charge to the $(3+1)$-dimensional Reissner-Nordstr\"{o}m AdS$_4$ black hole in the bulk, the so-called dyonic black hole \cite{Hartnoll:2007ai}. This background, which is a solution to the Einstein-Maxwell equations, is given by \cite{Romans:1991nq}
\begin{align}\label{eqt: dyon}
ds^2 &=\frac{r^2}{L^2}\Big(- f(r) dt^2 + dx^2 + dy^2\Big)+\frac{L^2}{r^2 f(r)} dr^2,\nonumber\\ 
A &=Q \left(\frac{r_0}{L^2}\right)\left(1-\frac{r_0}{r}\right) dt+ \frac{H}{2}\left(\frac{r_0}{L^2}\right)^2 \left(xdy-y dx\right)
\end{align}
with
\begin{align}\label{dyonicf}
f(r) = 1-\left(1+ Q^2+H^2\right) \Big(\frac{r_0}{r}\Big)^3+ \left(Q^2+H^2\right) \Big(\frac{r_0}{r}\Big)^4
\end{align}
where $r_0$ is the horizon radius given by the largest real root of $f(r_0)=0$,  and $H$ and $Q$ are two dimensionless parameters representing the magnetic and the electric charges of the black hole. We have chosen the gauge in which the spatial $\rm SO(2)$ symmetry is preserved.
The temperature of the black hole becomes
\begin{equation}
T = \frac{r_0}{4\pi L^2}\left(3-Q^2-H^2\right).
\end{equation}
The temperature vanishes in the extremal limit,  $Q^2+H^2= 3$, at which point the metric function $f(r)$ has a double zero.

The asymptotic form of the gauge field gives rise to both an electric charge density  and a homogeneous background magnetic field in the dual field theory \cite{Hartnoll:2007ai}. Denoting by $B$ and $\rho$ the background magnetic field and the electric charge density of the dual theory, respectively, one gets 
\bea
B=H\Big(\frac{r_0}{L^2}\Big)^2, \qquad \qquad \rho =\frac{2}{\kappa_4^2}\frac{r_0^2}{L^2}Q.
\eea
Also, as in the charged case, the asymptotic value of $A_t$ is identified in the boundary theory with the  chemical potential for the electric charge density $\mu=Q r_0/L^2$.

In the rest of the section, we work in the extremal limit, where the boundary field theory is at zero temperature\footnote{Note that in the extremal limit the dyonic black hole also has a finite entropy density $s$ whose expression is the same as given in (\ref{ece}). 
}. It is convenient to parametrize the extremal values of $H$ and $Q$ by
\bea
Q=\sqrt{3}~\cos\theta, \qquad\qquad H=\sqrt{3} ~\sin\theta.
\eea
Since the spatial $\rm SO(2)$ is unbroken, it is convenient to organize the fields such that they have definite $\rm SO(2)$ charge, as described previously.  To do so, we write $z=x+iy$. For our present purpose, we will be interested in the fluctuations $a_x,a_y,{h^x}_t,{h^y}_t$. These give rise to the charge-1 fields 
\beq
\az=\frac12(a_x-ia_y),\qquad\qquad  \hzt=\frac12({h^x}_t-i{h^y}_t)
\eeq
which satisfy the coupled Einstein-Maxwell equations
\begin{align}\label{eq:shearconouter}
&\wn u^4\hzt'(u)-4\alpha Hu^2f(u)\az'(u)+24\alpha Q \left(\wn\az(u) -\frac{H}{36\alpha }\hzt(u)\right)=0,\\
&u^4f(u)\hzt''(u)+ 4u^3 f(u)\hzt'(u)+ 24\alpha Qf(u)\az'(u)
+\frac{144\alpha H}{u^2} \left(\wn\az(u) -\frac{H}{36\alpha }\hzt(u)\right)=0,\\
&u^2f(u)\az''(u)+(uf'(u)+2f(u))u\az'(u)+\frac{1}{2\alpha Q}\hzt'(u)+
\frac{36\wn}{u^2f(u)} \left(\wn\az(u) -\frac{H}{36\alpha }\hzt(u)\right)=0.
\end{align}
These equations of course are not all independent, and essentially constitute a first and a second order differential equation for $\az(r),\hzt(r)$. Similar equations are satisfied by the charge $-1$ modes $\azb, \hzbt$. In fact, the equations for those fields are the same as those given above, with the simple change $\theta\mapsto -\theta$ (that is, $H\mapsto -H$).

The above system of equations may be solved by noting that they decouple into a single second order equation for
\begin{align}
\gz(u)&=\wn\az(u)-\frac{H}{36\alpha }\hzt(u),\\
\gzb(u)&=\wn\azb(u)+\frac{H}{36\alpha }\hzbt(u),
\end{align}
which are the components of the boundary electric field.
We find
\begin{align}\label{gammaeq}
&\frac{1}{4u^2f(u)}\Big[ 48H^2f(u)^2+24HQu^3f(u)[2f(u)-uf'(u)]\wn-4f(u)(H^2+3)(6\wn u)^2+(6\wn u)^4\Big]\gz(u)\nonumber\\
&+\Big[9u^2\wn^2[2f(u)+uf'(u)]-4H^2f(u)^2\Big]u^3\gz'(u)+\Big[9u^2\wn^2-H^2f(u)\Big]u^4f(u)\gz''(u)=0.
\end{align}
Given the appropriate solution to this equation, we then deduce $\az$ and $\hzt$ by solving the constraint equation, which takes the form
\beq\label{constraint}
\az'(u)=\frac{3}{u^2}\frac{3\wn u^4\gz'(u)-2QH \gz(u)}{9u^2\wn^2-H^2f(u)}.
\eeq
We note a subtlety in this expression, because the denominator may be singular. In fact, this denominator is clearly also present in (\ref{gammaeq}). Since we are interested in the small $\wn$ expansion, we will be lead to require 
\beq
u \ll \frac{H}{3\wn},
\eeq
and making this assumption, we will be able to extract the Hall conductivity.
The presence of the magnetic field apparently introduces a cutoff in the UV, but for sufficiently small $\wn$, we can get arbitrarily close to $u\to\infty$. Ultimately, this is tied to the fact that the $H\to 0$ and $\omega\to 0$ limits do not commute. Similar statements have appeared in studies of magnetohydrodynamics \cite{Hartnoll:2007ai, Hartnoll:2007ip, Hartnoll:2007ih, Buchbinder:2009mk}.

We proceed to solve by expanding in powers of $\omega$
\beq
\gz(u)=E^{(0)}_z(u)+\wn E^{(1)}_z(u)+\wn^2E^{(2)}_z(u)+\ldots~.
\eeq
The zeroth order equation
\beq
\left(u^6f(u)\frac{d^2}{du^2}+4u^5f(u)\frac{d}{du}-12\right)E^{(0)}_z(u)=0
\eeq
has the exact solution 
\begin{align}\label{gz01}
E^{(0)}_z(u)=c_1 f(u)&+c_2\Big\{\frac{43u^4-30u^3+396u^2-538u+201}{432 u^4(u-1)}\nonumber\\
&+\frac{f(u)}{2592}\Big[17\sqrt{2}\Big(\arctan\Big(\frac{u+1}{\sqrt{2}}\Big)-\frac{\pi}{2}\Big)+56\ln\Big(\frac{(u-1)^4}{u^4f(u)}\Big)\Big]\Big\}.
\end{align}
Asymptotically, this behaves as
\beq
E^{(0)}_z(u)|_{u\to\infty}\simeq c_1+(c_2-4c_1)\frac{1}{u^3}+\ldots~.
\eeq
This is a sensible asymptotic form, since at $\omega=0$, $\gz$ is a metric mode. In the interior, we find
\beq
E^{(0)}_z(u)|_{u\to1}\simeq
c_2\left[\frac{1}{6 (u-1)}+\frac19+\ldots\right]+c_1\Big[6(u-1)^2+\ldots\Big].
\eeq
Proceeding further in the $\omega$ expansion, we find at first order
\beq\label{gz1}
\left[u^6f(u)\frac{d^2}{du^2}+4u^5f(u)\frac{d}{du}-12\right]E^{(1)}_z(u)
+6\cot\theta\frac{u^3\left(uf'(u)-2f(u)\right)}{f(u)}E^{(0)}_z(u)=0.
\eeq
The general solution of this equation includes those solutions annihilated by the differential operator in square brackets (\ie, the same solutions found for $E^{(0)}_z(u)$) plus a particular solution. The two integration constants are associated with the former. Proceeding to all orders, the solution will take the form
\beq
\gz(u)=c_1(\omega)E^{(0)+}_z(u)+c_2(\omega)E^{(0)-}_z(u)+\sum_{n=1}^\infty\wn^n \hat E^{(n)}_z(u)
\eeq
where $E^{(0)\pm}_z(u)$ refer to the two independent solutions in (\ref{gz01}) and $\hat E^{(n)}_z(u)$ refer to the particular solutions at $n^{\rm th}$ order. The particular solution to equation (\ref{gz1}) can be obtained exactly, and has the asymptotic form\footnote{Although we do not need the following relations for our analysis of the Hall conductivity, we present them for the sake of completeness
\begin{align}
c_3&\equiv -\frac{1}{216}(13+17\sqrt{2}\pi)c_1+\frac{1}{2^8 3^7}\left(-11094 + 22253 \sqrt{2} \pi + (17\pi)^2\right)c_2,\nonumber\\
c_4&\equiv \frac{1}{54}(13+17\sqrt{2}\pi)c_1-\frac{1}{2^6 3^7}\left(-19518 +11237 \sqrt{2} \pi + (17\pi)^2\right)c_2.\nonumber
\end{align}}
\beq
\hat E^{(1)}_z(u)|_{u\to\infty}=\cot\theta\Big[c_3-6\frac{c_1}{u}+\frac{c_4}{u^3}+\ldots\Big].
\eeq

To apply the infalling boundary condition, we proceed to the near horizon geometry
\begin{align}
ds^2&=\frac{L^2}{6\eta^2}\Big(-dt^2+d\eta^2\Big)+\frac{r_0^2}{L^2}\Big(dx^2+dy^2\Big)\\
A&=\frac{Q}{6\eta}dt+\frac{Hr_0^2}{2L^4}(xdy-ydx)
\end{align}
where $\eta$ is defined as in \eqref{definition eta}. The Einstein-Maxwell equations can be decoupled in precisely the same way, and we find
\newcommand\et{\zeta}
\beq
\et^2(\et^2-2\sin^2\theta)\gz''(\et)-4\et\sin^2\theta\gz'(\et)+\left[ (\et^2-2)(\et^2-2\sin^2\theta)-2\et\sin2\theta\right]\gz(\et)=0
\eeq
where we have defined $\et=\omega\eta$. Remarkably, this has the exact solutions
\beq
\gz(\et)=c_+ e^{i\et}\left( 1+\frac{i}{\et}+\frac{i\sin\theta\ e^{i\theta}}{\et^2}\right)+
c_- e^{-i\et}\left( 1-\frac{i}{\et}-\frac{i\sin\theta\ e^{-i\theta}}{\et^2}\right).
\eeq
The infalling wave at $\et\to \infty$ is obtained by setting $c_-=0$.

Now the prescription of \cite{Faulkner:2009wj}  is that we should expand this solution for $\et\to 0$ (with $\wn/\zeta\to 0$) and match it to $E^{(0)}_z(u)$. We find
\begin{align}
\gz(\et)&\simeq \frac{i}{\et^2}\left(c_+e^{i\theta}-c_-e^{-i\theta}\right)\left[\sin\theta\Big(1-\frac12\et^2+\ldots\Big)+\et\cos\theta\Big(1+\frac12\et^2+\ldots\Big)\right]\nonumber\\
&+\et\left(c_+e^{i\theta}+c_-e^{-i\theta}\right)\left[\sin\theta\Big(\frac23-\frac{2}{15}\et^2+\ldots\Big)+\et\cos\theta\Big(-\frac13+\frac{1}{30}\et^2+\ldots\Big)\right].
\end{align}
We have written the solution in this form to highlight the fact that it smoothly interpolates between predominantly metric and gauge field as we vary $\theta$. In particular, the case of the charged black hole is of course obtained in the limit $\theta\to0$. We thus find the matching relations
\begin{align}
c_1&=\frac{i}{6\alpha^2\omega^2}c_+e^{i\theta}\sin\theta,\\
c_2&=4\alpha\omega c_+e^{i\theta}\sin\theta.
\end{align}

Finally, we solve the constraint (\ref{constraint}) to leading order in $\wn$ to obtain
\begin{align}
\az(u)|_{u\to\infty}&=c_5-6\cot \theta\frac{c_1}{u}+O(u^{-3}),\\
\hzt(u)|_{u\to\infty}&=-\frac{36\alpha}{H}(c_1-\wn c_5+\wn\cot\theta c_3)+O(u^{-3}).
\end{align}
Hence, we read off that to lowest order in $\wn$ 
\beq
c_1=-\frac{H}{36\alpha}\hzt(\infty)+\wn \az(\infty)=\gz(\infty).
\eeq
Similarly, denoting the constants involved in $\azb$ and $\hzbt$ by tildes, we have
\beq
\tilde c_1= \frac{H}{36\alpha}\hzbt(\infty)+\wn \azb(\infty)=\gzb(\infty),
\eeq
where we have taken the fields to be functions of $\omega$. This result (which should be noted does not depend on the details of the horizon boundary conditions) determines the Hall conductivity. 
The renormalized action read
\begin{align}
S_{\rm ren}&=\lim_{u\to\infty}\frac{2r_0}{9\kappa_4^2\alpha^2}\int dt~d^{2}x\Big[ - \hzt\hzbt+\frac32 \alpha Q(\az\hzbt+\azb\hzt)\nonumber\\
 &\phantom{\frac{1}{2}} \qquad \qquad \qquad \qquad \qquad  +\frac{u^4}{16}(\hzt\hzbt'+\hzbt\hzt')-9\alpha^2 u^2(\az\azb'+\azb\az')\Big].
\end{align}
The on-shell action then becomes
\begin{align}
S_{\rm on-shell}&= \frac{4L^{2}\cot\theta }{\kappa_4^2 }\int d^{2}x~d\omega\left[\omega\az(\omega) \azb(-\omega)\right]+\ldots, \nonumber\\
&= \frac{\rho}{B}\int d^{2}x~d\omega\left[2\omega\az(\omega) \azb(-\omega)\right] + \ldots~.\label{finalhall}
\end{align}
Note that 
\beq
\omega\az(\omega)\azb(-\omega)=\left[ \omega a_x(\omega)a_x(-\omega)+\omega a_y(\omega)a_y(-\omega)\right]-2i\left[ \omega a_x(\omega)a_y(-\omega)\right].
\eeq
The reality of the coefficient in (\ref{finalhall}) indicates that this result represents $\sigma_{xy}=\rho/B$ and $\sigma_{xx}=\sigma_{yy}=0$.

\section{Summary and Discussion}\label{sectionseven}

In this paper we have computed the shear viscosity and the conductivity of a (2+1)-dimensional boundary field theory at zero temperature and finite charge density, which is dual to the extremal Reissner-Nordstr\"{o}m $\ads_4$ black hole. A peculiar property of the boundary field theory is its finite entropy density at zero temperature, which makes one wonder about the existence of the hydrodynamics modes at small frequency. Taking the small frequency limit is subtle as the $\omega\to 0$ and $T\to 0$ limits do not generically commute, which makes the extrapolations to zero temperature of a transport coefficient obtained in a finite temperature system questionable.  Taking this subtlety into account, we showed that the ratio of the shear viscosity to the entropy ratio equals $1/4\pi$ in the zero-temperature boundary field theory. We have explicitly checked that the behavior of the two aforementioned transport coefficients does not change in higher dimensional analog of the (2+1)-dimensional boundary field theory, namely, independent of the dimension of the boundary field theory, the ratio of the shear viscosity to the entropy density takes the universal value of $1/4\pi$, saturating the KSS bound \cite{Kovtun:2004de},  whereas the conductivity has a power law behavior near $\omega\to 0$, along with the Drude $\delta$-function peak.

We computed the viscosity and conductivity using the appropriate Kubo formulas. It would be interesting to first find the corresponding diffusion constants from the poles of the retarded Green functions and then use hydrodynamics equations to extract the values of viscosity and conductivity. This approach is more involved as it requires taking the momentum to be small but non-vanishing, in which case one needs to do a double expansion in $\omega$ and $k^2$ in both the inner and outer regions. 

Our results for the ratio of the shear viscosity to the entropy density enlarges the domain of universality of the ratio to include theories dual to extremal black holes. The general argument given in \cite{Benincasa:2006fu} for the universality of the ratio in theories with finite chemical potential, or even  more general universality argument of \cite{Iqbal:2008by} are applicable only to backgrounds at finite (non-vanishing) temperature. Although we do not provide a rigorous argument for the universality of the ratio in theories dual to extremal black holes, our examples give evidence that, at least in the class of zero-temperature field theories dual to the extremal Reissner-Nordstr\"{o}m AdS$_{d+1}$ black holes, the ratio takes the universal value of $1/4\pi$. We believe that given the results of this paper, one should be able to extend the arguments of \cite{Iqbal:2008by, Benincasa:2006fu} to the cases where the background is extremal. 

We have shown that the longitudinal conductivity $\sigma(\omega/T=\infty,\vec k=0,\omega\to 0)$ scales as $\omega^2$. It is easy to understand this scaling behavior from our AdS/CFT calculations as a dimension 2 uncharged scalar operator in the IR CFT is responsible for such behavior in the boundary field theory. It would be interesting to understand this scaling behavior from field theory perspectives, and investigate its  possible relevance for quantum critical points. 

The Hall conductivity, on the other hand, is not determined by properties of the IR CFT, as  it follows from simple (Lorentz) symmetry arguments. We have included the calculation in this paper simply to show that the zero temperature limit can be taken. Moreover the analysis can be easily applied when fields are mixing in the case of finite $\vec k$. 

Shear viscosity and conductivity are just two examples of transport coefficients. Indeed,  one can also compute other transport coefficients at zero temperature similar to what we performed in this paper. An example would be the diffusion constant of a probe particle moving in the zero-temperature boundary field theory.     Finally, it would be interesting to compute the transport coefficients of the boundary theories dual to extremal non-relativistic backgrounds studied in \cite{Imeroni:2009cs,Adams:2009dm}.

\section*{Acknowledgments}

We would like to thank E. Fradkin, P. Phillips and J. Polchinski for helpful discussions.  R.G.L.\ and M.E.\ are supported by DOE grant FG02-91-ER40709 and J.I.J. is supported by a Fulbright-CONICYT fellowship.

\providecommand{\href}[2]{#2}\begingroup\raggedright\endgroup

\end{document}